# On the electromagnetic constitutive laws that are equivalent to spacetime metrics

D. H. Delphenich
Spring Valley, OH 45370 USA

**Abstract:** The "raising of both indices" in the components of the Minkowski electromagnetic field strength 2-form $F$ to give the components of the electromagnetic excitation bivector field $\mathfrak{H}$ can be regarded as being equivalent to an electromagnetic constitutive law, as well as being defined by the components of the spacetime metric. This notion is clarified, and the nature of the equivalent dielectric tensors and magnetic permeability tensors that are defined by some common spacetime metrics is discussed. The relationship of the basic construction to effective metrics is discussed, and, in particular, the fact that this effective metric is more general than the Gordon metric.

**1. Introduction.** – A new approach to classical electromagnetism has been emerging since the work of Kottler, Tamm, Cartan, and van Dantzig in the early Twentieth Century that goes by the name of "pre-metric electromagnetism" ([1]). The origin of the approach is the idea that the only place in which the background metric of spacetime enters into Maxwell's equations is by way of the Hodge * operator that allows one to define the codifferential $\delta F$ of the electromagnetic field strength 2-form $F$.

Some of the illusion that Maxwell's equations depends upon the Lorentzian structure is created by the fact that so many of the formulations of Maxwell's equations in terms of differential forms never go beyond the limitations of the "classical electromagnetic vacuum," whose electromagnetic constitutive properties can be summarized by a dielectric constant $\varepsilon_0$ and a magnetic permeability constant $\mu_0$. Indeed, they are often combined into the speed of light *in vacuo* $c = (\varepsilon_0 \mu_0)^{-1/2}$, which is almost as often set equal to unity. Hence, the fact that one has made a very restrictive assumption about the nature of the electromagnetic constitutive properties of the electromagnetic vacuum is made entirely invisible to the further analysis.

It is only when one considers Maxwell's equations in non-vacuous media that it becomes clear that one must re-introduce the constitutive properties of the medium $M$. In fact, it was observed by the founders of pre-metric electromagnetism that the role of the Lorentzian metric – i.e., defining the isomorphism $*: \Lambda^2 M \to \Lambda^2 M$ that maps $F$ to its dual – could be replaced by the introduction of an electromagnetic constitutive law $C: \Lambda^2 M \to \Lambda_2 M$ that mapped $F$ to the electromagnetic excitation bivector field $\mathfrak{H}$. The substitute for the * isomorphism then came about when one composed $C$ with the Poincaré duality isomorphism $\#: \Lambda_2 M \to \Lambda^2 M$ that resulted from introducing a volume element upon $M$. Indeed, the assumption that $C_x: \Lambda^2_x \to \Lambda_{2,x}$ is a linear isomorphism of fibers to fibers is a restriction of scope regarding the constitutive properties of the medium, since that amounts to assuming that the medium is non-dispersive (in the first sense of dispersion) and linear in its response to the imposition of electric and magnetic fields.

---

([1]) Since it is not the purpose of the present study to expose that approach to electromagnetism in all of its details, we refer the curious to the texts of Hehl and Obukhov [**1**] and the author himself [**2**], which include references to the original literature.

The way that one gets back to the Lorentzian metric on spacetime by starting with $C$ is by looking at the dispersion relation (in the second sense of dispersion) for "wave-like solutions" to the Maxwell equations. As long as one restricts oneself to linear media, it will be sufficient to consider plane-wave solutions, but for nonlinear media, it becomes necessary to realize that a given class of wave-like solutions will generally have its own dispersion law. Even in the case of linear media, one generally gets a dispersion law that is defined by a homogeneous, quartic polynomial $\mathfrak{D}(k)$ in the frequency-wave number 1-form $k = -\omega\, dt + k_i\, dx^i$. It is only in more restrictive cases that the quartic polynomial factors into a product of quadratic polynomials of Lorentzian type (viz., $-\omega^2 + k_1^2 + k_2^2 + k_3^2$), and in even more restrictive cases that the product of quadratic polynomials degenerates to the square of a Lorentzian polynomial. That is the point at which one can reconnect with general relativity.

It is interesting that, in a sense, the path from electromagnetism to general relativity also retraces the historical progression that Einstein made from the electrodynamics of moving bodies to a theory of gravitation. In effect, the theories of electromagnetism and gravitation were already "unified" in a different sense than what he was looking for later on. Namely, the overlap between the theories was defined by the light cones, which simultaneously governed the propagation of electromagnetic waves and defined the basis for the spacetime metric whose curvature explained the presence of gravitation.

However, there are other issues that come about as a result of that interface between electromagnetism and gravitation. For example:

1. Electromagnetic fields can have energy density, momentum flux, and stresses that can serve as the source of gravitational fields.

2. Gravitational fields can affect the propagation of light.

3. The electromagnetic constitutive properties of a medium can contribute to the metric directly as in the method of "effective metrics."

The first issue was first addressed by Rainich [3] and served as the basis of Wheeler's "already-unified field theory" [4]. We shall not discuss that topic further in the present study, however.

The second issue represents the expanding field of gravitational optics ([1]). Some of the optical phenomena that have come about as a result of the curvature of spacetime by gravitating bodies are redshifts (blueshifts, resp.) of the wave lengths of photons and the bending of light rays by gravitating bodies. Indeed, the latter phenomenon, which was observed during a solar eclipse in the early years of general relativity, has given rise to the astronomical phenomenon of "gravitational lensing," which is used to infer the presence of black holes. Furthermore, as was discussed by Plebanski [6], there is also a possible rotation of the plane of polarization of an electromagnetic wave by the gravitational field of a rotating mass distribution.

---

([1]) Since that field is already quite vast in scope, we shall simply refer the ambitious readers to the excellent modern monograph by Volker Perlick [5]. Some of the aspects of the construction that will be described here were also introduced by Jerzy Plebanski in 1960 [6] in the context of the scattering of plane electromagnetic waves by the gravitational field of an isolated physical system.

The concept of an effective (or "optical") metric for an electromagnetic medium in a state of relative motion to an observer goes back to Walter Gordon [**7**] in 1923, who only considered the case of a linear, isotropic, material medium that is in a state of motion relative to an observer. It was later made more rigorous by Pham Mau Quan [**8**] and Jürgen Ehlers [**9**]. The properties of the medium were extended to absorptive media by Chen and Kantowksi [**10**], and the possibility of using effective metrics as analogue models for spacetimes − such as Schwarzschild spacetimes – was discussed by Novello, *et al*. [**11, 12**]. Ulf Leonhardt examined the effective metrics of quantum dielectrics in [**13**] and expanded the application of effective metrics into the field of "transformation optics [**14**]," which relates to the propagation of electromagnetic waves in metamaterials and the possibility of "cloaking devices."

In the next section of this article, we shall discuss the topic of electromagnetic constitutive laws in a general mathematical and physical context. In the third section, we shall suggest a straightforward way of associating an electromagnetic constitutive law that is defined by a choice of spacetime metric. Basically, one treats the linear isomorphism of "raising both indices" of the components $F_{\mu\nu}$ to obtain $F^{\mu\nu}$ as being a map that is proportional to a linear isomorphism that turns $F_{\mu\nu}$ into $\mathfrak{H}^{\mu\nu}$. In the fourth section, we shall apply that association to various common metrics that one considers in general relativity, and in the fifth section, we show that the inversion of our association of a constitutive law with a metric will produce an effective metric that is more general in scope than the ones that were considered by Gordon, Quan, *et al*. Finally, in the last section, we will discuss some further issues that concern the equivalence in question.

The novelty of the present study is basically in the generalization of the association of Lorentzian metrics with linear, non-dispersive, electromagnetic constitutive laws and the examination of some examples of constitutive laws that are equivalent to various common metrics of general relativity.

**2. Electromagnetic constitutive laws.** – The two fundamental tensor fields of Maxwellian electromagnetism (at least, in its later, relativistic formulation) are the *Minkowski electromagnetic field strength 2-form F* and the *electromagnetic excitation bivector field* $\mathfrak{H}$. In the natural frame field $\{\partial_\mu = \partial / \partial x^\mu, \mu = 0, 1, 2, 3\}$ ([1]) and its reciprocal natural coframe field $\{dx^\mu, \mu = 0, 1, 2, 3\}$ that are defined by a local coordinate chart $\{U, x^\mu\}$, one can express these tensor fields in component form as:

$$F = \tfrac{1}{2} F_{\mu\nu} dx^\mu \wedge dx^\nu, \qquad \mathfrak{H} = \tfrac{1}{2} \mathfrak{H}^{\mu\nu} \partial_\mu \wedge \partial_\nu, \qquad (2.1)$$

in which $F_{\mu\nu}$ and $\mathfrak{H}^{\mu\nu}$ are anti-symmetric component matrices.

When one has a time + space splitting of the spacetime manifold $M$ (i.e., its tangent and cotangent bundles, $T(M)$ and $T^*(M)$, resp.), such that one identifies $\partial_0$ and $dx^0$ as the generators of the time lines in $T(M)$ and $T^*(M)$, resp., while the 3-frame $\{\partial_i, i = 1, 2, 3\}$ and the 3-coframe $\{dx^i, i = 1, 2, 3\}$ span the spatial sub-bundles, one can express the two fundamental fields in the forms:

$$F = dx^0 \wedge E - B, \qquad \mathfrak{H} = \partial_0 \wedge \mathbf{D} + \mathbf{H}, \qquad (2.2)$$

in which the spatial 1-form and vector field:

$$E = E_i \, dx^i \qquad \text{and} \qquad \mathbf{D} = D^i \, \partial_i \qquad (2.3)$$

are what usually get referred to as the *electric field strength* 1-form and *electric excitation* (or *displacement*) vector field, resp., while the spatial 2-form and bivector field:

$$B = \tfrac{1}{2} B_{ij} \, dx^i \wedge dx^j \qquad \text{and} \qquad H = \tfrac{1}{2} H^{ij} \, \partial_i \wedge \partial_j \qquad (2.4)$$

are related to the *magnetic excitation (or induction) vector field* **B** and the *magnetic field strength 1-form H*, resp., as spatial duals:

$$B_{ij} = \varepsilon_{ijk} B^k, \qquad H^{ij} = \varepsilon^{ijk} H_k. \qquad (2.5)$$

---

([1]) We shall define $x^0 = c_0 t$, where $c_0$ is the speed of light *in vacuo*. However, since we will be dealing the propagation of electromagnetic waves in a more general vacuum, we shall not introduce the usual simplification of setting $c_0$ equal to unity.

The physical relationship between $F$ and $\mathfrak{H}$ is potentially quite involved, especially when one looks at electromagnetic fields in material media ([1]). At the essence of the relationship is the formation of electric and magnetic dipoles as a result of the presence of the field $F$; i.e., the medium becomes *polarized*. For us, the field $\mathfrak{H}$ will represent the response of the medium to the presence of electric and magnetic fields, even though it actually consists of an excitation (**D**) and a field strength ($H$), while $F$ is composed of the opposite cases.

In the most general context, one has a map $C: \Lambda^2(M) \to \Lambda_2(M)$, $F \mapsto C(F)$ from the infinite-dimensional vector space $\Lambda^2(M)$ of 2-forms to the infinite-dimensional vector space $\Lambda_2(M)$ of bivector fields that we shall call the *electromagnetic constitutive law*; i.e.:

$$\mathfrak{H} = C(F). \qquad (2.6)$$

This map $C$ might be composed of an integral operator, a differential operator, and an algebraic one in some combination. The first possibility is referred to as *dispersion* (in the first sense of the term), and corresponds to the fact that the responses of most material media to the imposition of electric and magnetic fields are not perfectly local in time and space, but will depend upon past values (i.e., temporal dispersion) and spatially-neighboring values (i.e., spatial dispersion) of the electric and magnetic fields. Dispersion will also have the effect of making the Fourier transform of $\mathfrak{H}$ be a function of not only the Fourier transform of $F$, but also of frequency and wave numbers.

---

([1]) Good discussions of the issues associated with electromagnetic constitutive laws can be found in various books, such as Landau, Lifschitz, and Pitaevski [**15**], Post [**16**], Lindell [**17**], Hehl and Obukhov [**1**], and the author's book [**2**].

Constitutive laws that are defined by differential operators are not as common in physical applications, but we include them here only for the sake of completeness.

The algebraic part of $C$ is what usually gets the most attention in applications, which then corresponds to considering non-dispersive media, or at least ignoring what dispersion that there is. Since a purely algebraic operator is also purely local, one can characterize such a map $C$ by its restrictions $C_x : \Lambda^2_x \to \Lambda_{2,x}$ to the individual fibers of $\Lambda^2(M)$ and write the constitutive law in the local form:

$$\mathfrak{H}^{\mu\nu} = C^{\mu\nu}(x^\kappa, F_{\kappa\lambda}). \tag{2.7}$$

One can also put this relationship into the form:

$$\mathfrak{H}^{\mu\nu} = C^{\mu\nu\kappa\lambda}(x, F) F_{\kappa\lambda} + \mathfrak{H}^{\mu\nu}(x, 0), \tag{2.8}$$

in which $\mathfrak{H}^{\mu\nu}(x, 0)$ is the *zero-point field*; i.e., the residual state of polarization that is present when no electromagnetic field is applied, as one finds in ferromagnetism and ferro-electricity.

In the algebraic case, one usually assumes that $C$ is invertible. The first big distinction to be made regarding the purely algebraic constitutive laws is between linear and nonlinear ones. As usual, although linearity is the most commonly discussed case, it is always a small-amplitude approximation, at least in material media. However, even when one is discussing the electromagnetic vacuum, one might regard the onset of vacuum polarization when the electric or magnetic field strengths approach the Schwinger critical values as being the linear limit of the classical approximation. If the constitutive law is linear then the relationship (2.8) can be given the form:

$$\mathfrak{H}^{\mu\nu} = C^{\mu\nu\kappa\lambda}(x) F_{\kappa\lambda} + \mathfrak{H}^{\mu\nu}(x, 0), \tag{2.9}$$

and in the absence of a zero-point field:

$$\mathfrak{H}^{\mu\nu} = C^{\mu\nu\kappa\lambda}(x) F_{\kappa\lambda}. \tag{2.10}$$

For linear, non-dispersive constitutive laws, one can also express this relationship in time + space form using a 6×6 real matrix $C^{IJ}$, $I, J = 1, \ldots, 6$ that is composed of 3×3 blocks. This goes back to the fact that the fibers of both $\Lambda^2$ and $\Lambda_2$ are six-dimensional, real, vector spaces, which then admit frames with six members. A particularly useful choice in both cases is $\{b^I, I = 1, \ldots, 6\}$ and $\{\mathbf{b}_I, I = 1, \ldots 6\}$, respectively, in which one defines:

$$\mathbf{b}_i = \partial_0 \wedge \partial_i, \qquad \mathbf{b}^i = \tfrac{1}{2}\varepsilon^{ijk} \partial_j \wedge \partial_k, \tag{2.11}$$

$$b^i = dx^0 \wedge dx^i, \qquad b_i = \tfrac{1}{2}\varepsilon_{ijk} dx^i \wedge dx^j, \tag{2.12}$$

so one has:

$$F = E_i b^i - B^i b_i, \qquad \mathfrak{H} = D^i \mathbf{b}_i + H_i \mathbf{b}^i. \tag{2.13}$$

If one writes the components of both $F$ and $\mathfrak{H}$ as column vectors then the relationship (2.10) can be written in block-matrix form as:

$$\begin{bmatrix} D^i \\ H_i \end{bmatrix} = \begin{bmatrix} \varepsilon^{ij} & -\alpha^i_j \\ \beta^j_i & -\tilde{\mu}_{ij} \end{bmatrix} \begin{bmatrix} E_j \\ -B^j \end{bmatrix}; \qquad (2.14)$$

i.e.:

$$D^i = \varepsilon^{ij} E_j + \alpha^i_j B^j, \qquad H_i = \beta^j_i E_j + \tilde{\mu}_{ij} B^j. \qquad (2.15)$$

Tamm [**18**] refers to these equations as the *Minkowski equations*.

One then sees that $\varepsilon^{ij}$ is the component matrix of the *dielectric strength tensor* and $\tilde{\mu}_{ij}$ is the inverse of the component matrix of the *magnetic permeability tensor*, while the matrices $\alpha^i_j$ and $\beta^i_j$ generally vanish, except when one has such phenomena as Faraday rotation and optical activity to contend with. As it will turn out, most of the constitutive laws that we shall be concerned with in this study will have vanishing $\alpha$ and $\beta$ matrices.

Typically, one assumes that both $\varepsilon^{ij}$ and $\tilde{\mu}_{ij}$ are symmetric, invertible matrices. Hence, there will be a special spatial frame (viz., an eigenvector frame) for each matrix in which they become diagonal matrices with non-zero diagonal elements. Such frames are referred to as *principal axis* frames, while the diagonal elements are principal values of the dielectric strengths and magnetic permeabilities. If the $\alpha$ and $\beta$ matrices vanish then one will find that $C^{IJ}$ will be a diagonalizable matrix, although the two principal axes that diagonalize $\varepsilon^{ij}$ and $\tilde{\mu}_{ij}$ do not generally have to coincide, since one is, after all, dealing with a six-dimensional space, not a three-dimensional one.

Two further reductions are common in practice: First of all, if the components $C^{IJ}(x)$ are constant functions of $x$ then the medium will be *homogeneous*, or one can also specify that it is electrically homogeneous when $\varepsilon^{ij}$ are constants or magnetically homogeneous when $\tilde{\mu}_{ij}$ are constants. One sees that the concept of homogeneity is unavoidably frame-dependent, and thus not susceptible to a relativistic formulation ([1]).

A second reduction relates to the possibility that the medium is *isotropic* in its electric or magnetic properties, or possibly both. In such a case, the principal values of the $\varepsilon^{ij}$ and $\tilde{\mu}_{ij}$ matrices will all be equal.

The extreme case of an electromagnetic constitutive law is the one that is most commonly used in relativity theory, namely, the *classical vacuum*, for which:

$$[C] = \begin{bmatrix} \varepsilon_0 \delta^{ij} & 0 \\ 0 & -\dfrac{1}{\mu_0} \delta_{ij} \end{bmatrix}. \qquad (2.16)$$

---

([1]) Of course, the concept of a Lorentz transformation does not become relevant until after one has obtained the dispersion law (in the second sense of dispersion) that is associated with $C$ and then only if it reduces to the square of a quadratic law.

Hence, it is non-dispersive, linear, isotropic, and homogeneous.

When one formulates the theory of electromagnetism in this form, it suddenly seems harder to justify that this formulation implicitly contains such a manifestly non-relativistic concept as homogeneity at its basis. That is, perhaps, one of the drawbacks to using the popular system of "natural" units, which makes $c_0$ invisible to the equations, since, after all, one also has:

$$c_0 = \frac{1}{\sqrt{\varepsilon_0 \mu_0}}, \tag{2.17}$$

and in the eyes of pre-metric electromagnetism, the electromagnetic constitutive properties of spacetime are more fundamental than its metric, which is a consequence of the assumptions that one makes about $C$. Thus, the idea that $\varepsilon_0$ and $\mu_0$ should represent universal constants, and not just special cases of a more general situation, seems less than definitive at the fundamental level.

**3. Electromagnetic constitutive laws that are defined by spacetime metrics.** – In conventional relativity theory, one usually does not refer to the distinction between $F$ and $\mathfrak{H}$ as something that is based in the response of the medium to the imposition of $F$, since conventional relativity usually assumes that the medium is the classical vacuum, which should not really have a response, except perhaps a rescaling of the components. Rather, the role of $\mathfrak{H}$ is played by the bivector field:

$$\mathbf{F} = \tfrac{1}{2} F^{\mu\nu} \partial_\mu \wedge \partial_\nu, \tag{3.1}$$

whose components are obtained by "raising the indices" of $F$ using the spacetime Lorentzian metric $g$, namely:

$$F^{\mu\nu} = g^{\mu\kappa} g^{\nu\lambda} F_{\kappa\lambda}. \tag{3.2}$$

Since the matrix $F_{\kappa\lambda}$ is antisymmetric, the double sum over $\kappa$ and $\lambda$ will take the form:

$$F^{\mu\nu} = \tfrac{1}{2}(g^{\mu\kappa} g^{\nu\lambda} - g^{\mu\lambda} g^{\nu\kappa}) F_{\kappa\lambda} = \tfrac{1}{2} g^{\mu\nu\kappa\lambda} F_{\kappa\lambda}, \tag{3.3}$$

in which we have defined:

$$g^{\mu\nu\kappa\lambda} = g^{\mu\kappa} g^{\nu\lambda} - g^{\mu\lambda} g^{\nu\kappa}. \tag{3.4}$$

One can express this relationship in time + space form by setting:

$$E^i = F^{0i} = \tfrac{1}{2} g^{0i\kappa\lambda} F_{\kappa\lambda} = g^{0i0j} F_{0j} + \tfrac{1}{2} g^{0ijk} F_{jk} = g^{0i\,0j} E_j + \tfrac{1}{2} g^{0ijk} B_{jk}, \tag{3.5}$$

$$B^{ij} = F^{ij} = \tfrac{1}{2} g^{ij\kappa\lambda} F_{\kappa\lambda} = g^{ij0k} F_{0k} + \tfrac{1}{2} g^{ijkl} F_{kl} = g^{ij\,0k} E_k + \tfrac{1}{2} g^{ijkl} B_{kl}, \tag{3.6}$$

and if we replace $B_{jk}$ with $\varepsilon_{jkl} B^l$ and $B^{ij}$ with $\varepsilon^{ijk} B_k$ then we will get:

$$E^i = a^{ij} E_j + b^i_j B^j, \qquad B_i = c^j_i E_j + d_{ij} B^j, \qquad (3.7)$$

in which:

$$a^{ij} = g^{0i0j} = g^{00} g^{ij} - g^{0i} g^{0j}, \qquad (3.8)$$
$$b^i_j = \tfrac{1}{2} g^{0ikl} \varepsilon_{klj} = \tfrac{1}{2} (g^{0k} g^{il} - g^{0l} g^{ik}) \varepsilon_{klj}, \qquad (3.9)$$
$$c^j_i = \tfrac{1}{2} g^{kl0j} \varepsilon_{kli} = \tfrac{1}{2} (g^{0k} g^{jl} - g^{0l} g^{kj}) \varepsilon_{kli}, \qquad (3.10)$$
$$d_{ij} = \tfrac{1}{4} g^{mnkl} \varepsilon_{mni} \varepsilon_{klj} = \tfrac{1}{4} (g^{mk} g^{nl} - g^{ml} g^{nk}) \varepsilon_{mni} \varepsilon_{klj}. \qquad (3.11)$$

In these calculations, we have taken advantage of the fact that the parenthetic expressions are anti-symmetric in $kl$, as are the $\varepsilon_{kli}$ and $\varepsilon_{klj}$.

Some symmetries in the submatrices are immediate from these definitions:

$$a^{ij} = a^{ji}, \qquad c^j_i = b^i_j, \qquad d_{ij} = d_{ji}. \qquad (3.12)$$

Hence, the matrix $[c]$ is simply the transpose of the matrix $[b]$.

Hence, if the 6×6 matrix that corresponds to $g^{\mu\nu\kappa\lambda}$ is:

$$G^{IJ} = \begin{bmatrix} a^{ij} & b^i_j \\ \hline b^j_i & d_{ij} \end{bmatrix} \qquad (3.13)$$

then one must have that $G^{IJ}$ is symmetric:

$$G^{IJ} = G^{JI}. \qquad (3.14)$$

The symmetries (3.12) in the submatrices will then give us a set of necessary conditions for an electromagnetic constitutive law to represent the raising of indices due to a Lorentzian metric. Naively, if the matrix $b^i_j$ is essentially arbitrary then those symmetries will reduce the total set of possibilities for an invertible 6×6 real matrix to represent $g^{\mu\nu\kappa\lambda}$ for some Lorentzian metric $g$ from 36 to 21. However, there are only ten independent parameters to the symmetric matrix $g_{\mu\nu}$, so we would require an additional 11 algebraic relations on the matrices $a^{ij}$, $b^i_j$, $d_{ij}$ in order to make the quadratic system of equations (3.8)-(3.11) soluble for $g_{\mu\nu}$ when one is given $a^{ij}$, $b^i_j$, $d_{ij}$.

One case in which the system can be solved directly is when the matrix $g^{ij}$ is invertible. If we call its inverse $\tilde{g}_{ij}$ then we can solve (3.9) for $g^{0k}$ immediately:

$$g^{0k} = \tfrac{1}{2} \varepsilon^{kij} \tilde{g}_{il} b^l_j. \qquad (3.15)$$

We can then solve (3.8) for $g^{00}$:

$$g^{00} = \tfrac{1}{3} \tilde{g}_{ik} (a^{ki} + g^{0k} g^{0i}). \qquad (3.16)$$

That means that there are only 6 free parameters in this solution, which come from the symmetric matrix $g^{ij}$. Hence, not all possible matrices $a^{ij}$, $b^i_j$, $d_{ij}$ can be obtained in this way.

If one considers the inverse $g_{\mu\nu}$ to the matrix $g^{\mu\nu}$ then by the definition of inverse, one should have:

$$\delta^i_j = g_{j\mu} g^{\mu i} = g^{0i} g_{0j} + g^{ik} g_{kj}. \tag{3.17}$$

Hence, if $g_{ij} = \tilde{g}_{ij}$ then that would have to imply the vanishing of $g^{0i}$. Thus, one will not generally have that $g_{ij}$ is the inverse of $g^{ij}$. However, that will be the case when $g$ is in time + space form:

$$g = g_{00} dt^2 + g_{ij} dx^i dx^j, \tag{3.18}$$

which will make:

$$g_{i0} = g_{0j} = 0. \tag{3.19}$$

This class of component matrices includes diagonal matrices, which itself includes many of the commonly-used space-time metrics, such as Robertson-Walker, Schwarzschild, de Sitter, and anti-de-Sitter. One sees that for metrics of the form (3.18), equations (3.8)-(3.11) will simplify to:

$$a^{ij} = g^{00} g^{ij}, \quad b^i_j = c^j_i = 0, \quad d_{ij} = g_{ij}. \tag{3.20}$$

Note that if one changes the signs of $g^{00}$ and $g^{ij}$ then $a^{ij}$ will keep the same sign, but $d_{ij}$ will change sign. Thus, the matrix:

$$G^{IJ} = \begin{bmatrix} g^{00} g^{ij} & 0 \\ \hline 0 & g_{ij} \end{bmatrix} \tag{3.21}$$

will be sensitive to the sign convention that one chooses for $g$.

In particular, for Minkowski space, if the metric is given the form:

$$g = \eta_{\mu\nu} dx^\mu dx^\nu = c_0^2 dt^2 - \delta_{ij} dx^i dx^j, \tag{3.22}$$

so:

$$g_{00} = c_0^2, \quad g_{ij} = -\delta_{ij}, \quad g^{00} = \frac{1}{c_0^2}, \quad g^{ij} = -\delta^{ij}, \tag{3.23}$$

then we will get:

$$G^{IJ} = \begin{bmatrix} -c_0^{-2} \delta^{ij} & 0 \\ \hline 0 & -\delta_{ij} \end{bmatrix}, \tag{3.24}$$

which says that:

$$E^i = -c_0^{-2} E_i, \quad B_i = -B^i. \tag{3.25}$$

If we had chosen the opposite sign convention for $g$ then the result would have been:

$$G^{IJ} = \begin{bmatrix} -c_0^{-2}\delta^{ij} & 0 \\ 0 & \delta_{ij} \end{bmatrix}, \tag{3.26}$$

which is not the same as (3.24). We shall find that the latter sign convention is most convenient for the purposes of electromagnetic constitutive laws, and is also the one that most researchers in general relativity seem to prefer.

In order to motivate our basic association of raising the indices of $F_{\mu\nu}$ with a linear, non-dispersive, electromagnetic constitutive law, we first note that if we replace $c_0^{-2}$ with $\varepsilon_0 \mu_0$ in (3.26) then we can factor $-\mu_0$ out of the right hand side and get:

$$G^{IJ} = -\mu_0 \begin{bmatrix} \varepsilon_0 \delta^{ij} & 0 \\ 0 & -\frac{1}{\mu_0}\delta_{ij} \end{bmatrix}. \tag{3.27}$$

In this form, the matrix that we have produced represents the electromagnetic constitutive law for a linear, non-dispersive, isotropic, homogeneous medium, that is proportional to the classical vacuum (2.16).

The basic association that we will now make is to regard:

$$F^{\mu\nu} = -\mu_0 \mathfrak{H}^{\mu\nu}. \tag{3.28}$$

Hence, the raising of the indices of $F_{\mu\nu}$ will be proportional to an electromagnetic constitutive law $C$:

$$g^{\kappa\lambda\mu\nu} = -\mu_0 C^{\kappa\lambda\mu\nu}. \qquad (G^{IJ} = -\mu_0 C^{IJ}). \tag{3.29}$$

Referring back to (3.8)-(3.11), that will then make:

$$\varepsilon^{ij} = -\frac{1}{\mu_0} a^{ij} = -\frac{1}{\mu_0}(g^{00} g^{ij} - g^{0i} g^{0j}) = \varepsilon^{ji}, \tag{3.30}$$

$$\alpha^i_j = -\frac{1}{\mu_0} b^i_j = -\frac{1}{2\mu_0}(g^{0k} g^{il} - g^{0l} g^{ik})\varepsilon_{klj} = \beta^j_i, \tag{3.31}$$

$$-\tilde{\mu}_{ij} = -\frac{1}{\mu_0} d_{ij} = -\frac{1}{4\mu_0}(g^{mk} g^{nl} - g^{ml} g^{nk})\varepsilon_{mni}\varepsilon_{klj} = -\tilde{\mu}_{ji}. \tag{3.32}$$

If an electromagnetic constitutive law $C$ can be obtained from a Lorentzian metric $g$ in this way then will say that $C$ is *equivalent to* the metric $g$. Note that we are using the vacuum magnetic permeability $\mu_0$ as a unit conversion constant in the same way that special relativity often uses $c_0$ as a way of converting from time units to distance units.

When $g_{\mu\nu}$ has the time + space form, equations (3.30)-(3.32) will reduce to:

$$\varepsilon^{ij} = -\frac{1}{\mu_0} g^{00} g^{ij}, \qquad \alpha^i_j = \beta^j_i = 0, \quad \tilde{\mu}_{ij} = \frac{1}{\mu_0} g_{ij}. \tag{3.33}$$

One can solve these equations for the components $g^{00}$, $g^{ij}$:

$$g^{00} = -\tfrac{1}{3}\mu_0^2 \varepsilon^{ij} \tilde{\mu}_{ij}, \qquad g^{ij} = \frac{1}{\mu_0} \mu^{ij} \quad (\text{so } g_{00} = -\frac{3}{\mu_0^2} \tilde{\varepsilon}_{ij} \mu^{ij}, \; g_{ij} = \mu_0 \tilde{\mu}_{ij}). \tag{3.34}$$

One easily verifies that when $\varepsilon^{ij} = \varepsilon_0 \, \delta^{ij}$ and $\mu^{ij} = \mu_0 \, \delta^{ij}$, one does, in fact, define the Minkowski space metric by way of (3.34).

Hence, when the component matrix of the space-time metric has the time + space form, the electromagnetic constitutive law that it defines will have the properties that:

1. $\varepsilon^{ij}$ will be proportional to $\mu^{ij}$, and the proportionality factor will be:

$$Z = -\frac{1}{\mu_0^2} g^{00} = \tfrac{1}{3} \varepsilon^{ij} \tilde{\mu}_{ij}. \tag{3.35}$$

When $\varepsilon^{ij}$ (and thus, $\tilde{\mu}_{ij}$) are of the form $\varepsilon \, \delta^{ij}$ ($1 / \mu \, \delta_{ij}$, resp.), Z will be become $\varepsilon / \mu$, whose absolute value is sometimes referred to as the *impedance* of the isotropic medium. (This condition on the dielectric and magnetic properties of the medium is also sometimes referred to as "impedance matching.")

2. Consequently, the principal axes for both $\varepsilon^{ij}$ and $\tilde{\mu}_{ij}$ will be the same.

3. The medium will exhibit no Faraday rotation or optical activity (since $\alpha^i_j = \beta^j_i = 0$).

4. The spatial geometry will be dictated by its magnetic properties (or, equivalently, its dielectric properties).

Note that the typical constitutive law for anisotropic optical media, for which $\varepsilon^{ij}$ is symmetric and positive-definite, but arbitrary, and $\tilde{\mu}_{ij} = (1/\mu) \, \delta_{ij}$, would not be included in this class of constitutive laws, since it would not satisfy the first requirement.

**4. Examples from general relativity.** We shall now see what sort of electromagnetic constitutive laws are defined by some of the common Lorentzian metrics that are considered by general relativity. (The forms that we shall use are found in Grøn and Hervik [**19**] and Landau and Lifschitz [**20**].)

*a. Isotropic metrics.* Many of the most popularly-studied metrics of general relativity describe spatially-isotropic space-times. An isotropic metric will generally take the form:

$$ds^2 = -c_0^2 a(t, r) dt^2 + b(t, r)[dx^2 + dy^2 + dz^2] \tag{4.1}$$

in Cartesian coordinates ($r^2 = x^2 + y^2 + z^2$) and:

$$ds^2 = -c_0^2 a(t, r) dt^2 + B(t, r) dr^2 + C(t, r) r^2 (d\theta^2 + \sin^2\theta \, d\phi^2) \tag{4.2}$$

in spherical coordinates.

Since $C(t, r)$ can be assumed to be non-zero (the contrary case would constitute a singularity of a sort), one can factor it out of the last two terms and express $ds^2$ in the form:

$$ds^2 = -c_0^2 a(t, r) dt^2 + \alpha(t, r) [\beta(t, r) dr^2 + r^2 d\Omega^2], \tag{4.3}$$

in which $\alpha(t, r)$, $\beta(t, r)$ are new functions. Hence, the bracketed term is conformal to the Euclidian metric in spherical coordinates iff $\beta(t, r) = 1$.

Since Cartesian coordinates are more convenient for the sake of dielectric strength and magnetic permeability tensors, but isotropic metrics are more often given in spherical coordinates, we shall first show how to relate the relevant components of the metric in spherical coordinates to the corresponding ones in Cartesian coordinates and then express the dielectric strength and magnetic permeability in general form.

Basically, one notes that since $dr^2 = dx^2 + dy^2 + dz^2$, one would expect that the coefficient of $dr^2$ in (4.1) should equal the coefficient of $dr^2$ in (4.3):

$$b = \alpha\beta. \tag{4.4}$$

In fact, that is all we need, since we can set:

$$g_{00} = -c_0^2 a(t, r), \qquad g_{ij} = b \, \delta_{ij} = \alpha\beta \, \delta_{ij}, \tag{4.5}$$

so:

$$g^{00} = -c_0^{-2} a(t, r)^{-1}, \qquad g^{ij} = \frac{1}{b} \delta^{ij} = \frac{1}{\alpha\beta} \delta^{ij}. \tag{4.6}$$

If one replaces $c_0^{-2}$ with $\varepsilon_0 \mu_0$ then one will see from (3.33) that the dielectric strength and magnetic permeability tensors (in Cartesian coordinates) will then take the general form:

$$\varepsilon^{ij}(t, r) = \varepsilon(t, r) \, \delta^{ij}, \qquad \mu^{ij}(t, r) = \mu(t, r) \, \delta^{ij}, \tag{4.7}$$

in which:

$$\varepsilon(t, r) = \frac{\varepsilon_0}{a(t,r)b(t,r)} = \frac{\mu_0}{a\alpha\beta}, \qquad \mu(t, r) = \frac{\mu_0}{b(t,r)} = \frac{\mu_0}{\alpha\beta}. \tag{4.8}$$

One can then also compute a speed of light in the medium and an index of refraction for it:

$$c(t, r) = \frac{1}{\sqrt{\varepsilon\mu}} = c_0 \sqrt{a}\, b = \sqrt{a}\, \frac{\alpha\beta}{\mu_0}, \tag{4.9}$$

$$n(t, r) = \frac{c_0}{c(t,r)} = \frac{1}{b\sqrt{a}} = \frac{\mu_0 c_0}{\alpha\beta\sqrt{a}}. \tag{4.10}$$

*b. Robertson-Walker metric.* The Robertson-Walker metric is really a class of solutions to the Einstein field equations that describe all spatially homogeneous and isotropic cosmological models. Such space-times then admit six Killing vector fields that correspond to the infinitesimal generators of the Lie group of rigid motions in three-dimensional Euclidian space.

The basic metric in spherical coordinates is:

$$ds^2 = -c_0^2\, dt^2 + a(t)^2\, [(1 - kr^2)^{-1}\, dr^2 + r^2\, (d\theta^2 + \sin^2\theta\, d\phi^2)], \tag{4.11}$$

which is:

$$ds^2 = -c_0^2\, dt^2 + \frac{a(t)^2}{1 - kr^2}\, [dx^2 + dy^2 + dz^2] \tag{4.12}$$

in Cartesian coordinates.

In these expressions, $k$ is a constant that relates to the scalar curvature of the spatial hypersurfaces. That is, when $k > 0$, those spatial hypersurfaces will have constant positive scalar curvature, and will be referred to as *closed* cosmological models. When $k = 0$, they will be flat, Euclidian spaces, and when $k < 0$, they will have constant negative scalar curvature, which are then referred to as *open* cosmological models. The determination of the sign of $k$ generally comes down to the estimating the average mass density of the universe, which is itself quite small at the cosmological scale ([1]), and has the effect that the scalar curvature of "space" is also quite close to zero ([2]), and thus its sign is rather unstable with respect to successive refinements of the estimate on the average mass density ([3]).

The specific form of the expansion factor $a(t)$ must be obtained by first specifying a form for the right-hand side of the Einstein equations – i.e., a specific form for the energy-momentum-stress tensor $T_{\mu\nu}$, which will then result in an ordinary differential equation for $a(t)$. A popular form was given by Friedmann, who assumed that the matter in the universe (at the cosmological scale of distance) was a perfect, barotropic fluid, with a mass density of $\rho$, a pressure (or tension, depending upon the sign) $p$, and a four-velocity $u_\mu$, which made:

$$T_{\mu\nu} = (\rho + p)\, u_\mu u_\mu + p\, g_{\mu\nu}, \quad p(\rho) = w\rho. \tag{4.13}$$

---

([1]) Currently, it is estimated to be $6.0 \times 10^{-27}$ kg / m$^3$.
([2]) The current estimate of the radius of curvature of "space" is 50 billion light-years, as compared to the distance to the initial "Big Bang" singularity, which is 13.8 billion light-years.
([3]) For that reason, after too many decades of listening to the latest estimates, one gets somewhat skeptical about whether the "open vs. closed" debate has been resolved with any finality.

The proportionality constant *w* can take various forms, such as – 1 for a Lorentz-invariant vacuum energy (LIVE), 0 for cosmic dust, and 4/3 for radiation.

The resulting ordinary differential equation is then the *Friedmann equation*, and we refer the curious to the literature on general relativity and cosmology for a more thorough discussion, since our main concern here is with the properties of the equivalent electromagnetic constitutive law.

From the general discussion in the previous subsection, we now have:

$$\varepsilon(t, r) = \frac{1 - kr^2}{a(t)^2} \varepsilon_0, \quad \mu(t, r) = \frac{1 - kr^2}{a(t)^2} \mu_0. \tag{4.14}$$

Hence, the Robertson-Walker electromagnetic vacuum behaves like the classical vacuum, insofar as it is linear, non-dispersive, and isotropic, but it is no longer homogeneous, since it varies in time and with distance from the observer.

This metric gives a constant space-time impedance as a result, namely, from (3.35):

$$Z = \frac{\varepsilon_0}{\mu_0}; \tag{4.15}$$

i.e., the classical value, and the speed of light is now:

$$c(t, r) = \frac{a(t)^2}{1 - kr^2} c_0. \tag{4.16}$$

This value will then vanish at infinity, diverge when $r = 1/\sqrt{k}$ (for $k > 0$), and vary in time as $a(t)^2$.

One can also obtain the index of refraction as:

$$n(t, r) = \frac{1 - kr^2}{a(t)^2}, \tag{4.17}$$

which will then have the opposite asymptotic properties to $c(t, r)$.

The main optical effect that is associated with the expanding-universe geometry that astronomers use is the redshift of wavelengths due to the expansion of distances, in general. One can also use the gravitational lensing of light rays by stellar objects, which we will mention in the next section, to estimate the Hubble constant.

*c. Schwarzschild metric.* The Schwarzschild metric describes the geometry of space in the neighborhood of most celestial bodies, and comes about from the assumption that the geometry of space-time is spherically-symmetric and static, which also describes the distribution of gravitating mass that serves as the source of the metric field. However, the Schwarzschild solution itself is valid only outside of that mass distribution.

The basic metric, in spherical coordinates, is:

$$g = -c_0^2 \left(1 - \frac{r_S}{r}\right) dt^2 + \left(1 - \frac{r_S}{r}\right)^{-1} dr^2 + r^2 d\theta^2 + r^2 \sin^2\theta \, d\phi^2, \tag{4.18}$$

in which $r_S$ is a characteristic distance that is called the *Schwarzschild radius:*

$$r_S = \frac{2Gm}{c_0^2}, \tag{4.19}$$

where $G$ is Newton's universal gravitational constant, and $m$ is the gravitating mass.

That radius can be characterized as the radius within which the mass of the body would have to be concentrated in order for the escape velocity to be speed of light. For the Earth, it amounts to about 9 mm, and for the Sun, it is about 3 km. Thus, its existence becomes somewhat moot for any celestial body whose density does not approach the critical density for the Schwarzschild radius to approach the actual radius, since once one goes inside of the gravitating mass distribution, one can no longer use the metric (4.18), which is an exterior solution.

Since this metric has the functional form of an isotropic metric, we now have:

$$\varepsilon(r) = \varepsilon_0, \qquad \mu(r) = \left(1 - \frac{r_S}{r}\right)\mu_0. \tag{4.20}$$

Thus, the dielectric constant of the vacuum is unchanged from the classical value, but the magnetic permeability goes to zero as one approaches the Schwarzschild radius and goes to the classical value as one gets infinitely distant from it.

This time, we see that the spatial impedance varies from the classical value according to the inverse of the Schwarzschild factor:

$$Z = \tfrac{1}{3}\varepsilon^{ij}\tilde{\mu}_{ij} = \frac{\varepsilon_0}{\mu_0}\left(1 - \frac{r_0}{r}\right)^{-1}. \tag{4.21}$$

Hence, it will approach the classical vacuum value as $r$ becomes infinite and diverge as $r$ approaches the Schwarzschild radius.

The speed of light now takes the form:

$$c(r) = \left(1 - \frac{r_S}{r}\right)^{-1/2} c_0, \tag{4.22}$$

which goes to infinity as $r$ approaches $r_S$, and goes to the classical value $c_0$ as $r$ becomes infinite.

The index of refraction that one gets is then:

$$n(r) = \sqrt{1 - \frac{r_S}{r}}. \tag{4.23}$$

The observed optical effects of the Schwarzschild metric include the redshift of wave lengths when one gets closer to the center of the gravitating body ([1]) and the gravitational lensing of celestial objects due to the bending of light rays in their vicinity. Indeed, the experimental observation of the bending of light rays by Sun during a solar eclipse is one of the most well-established experimental tests of general relativity.

*d. Reissner-Nordström metric.* The difference between the Reissner-Nordström metric and the Schwarzschild metric is based upon the fact that the latter is assumed to describe the gravitational field of a *charged*, spherically-symmetric, static mass distribution. The basic effect in terms of the metric components is to replace $1 - \frac{r_S}{r}$ with:

$$1 - \frac{r_S}{r} - \kappa \frac{q^2}{r^2}.$$

Thus:

$$\varepsilon(r) = \varepsilon_0, \qquad \mu(r) = \left(1 - \frac{r_S}{r} + \kappa \frac{q^2}{r^2}\right)\mu_0, \qquad (4.24)$$

so:

$$Z = \frac{\varepsilon_0}{\mu_0}\left(1 - \frac{r_0}{r} + \kappa \frac{q^2}{r^2}\right)^{-1}, \qquad c(r) = \left(1 - \frac{r_S}{r} + \kappa \frac{q^2}{r^2}\right)^{-1/2} c_0. \qquad (4.25)$$

This time, as $r$ approaches $r_S$, this term in parentheses will approach $-\kappa \frac{q^2}{r^2}$, and it will still go to unity as $r$ becomes infinite.

The index of refraction now takes the form:

$$n(r) = \sqrt{1 - \frac{r_S}{r} + \kappa \frac{q^2}{r^2}}. \qquad (4.26)$$

*e. LTB metrics.* The Lemaître-Tolman-Bondi metrics allow one to introduce spatial inhomogeneity, while preserving isotropy. In spherical coordinates, they take the general form:

$$ds^2 = -c_0^2 dt^2 + X(t,r)^2 dr^2 + R(t,r)^2 d\Omega^2, \qquad (4.27)$$

or:

$$ds^2 = -c_0^2 dt^2 + \left(\frac{R(t,r)}{r}\right)^2\left[\left(\frac{r X(t,r)}{R(t,r)}\right)^2 dr^2 + r^2 d\Omega^2\right]. \qquad (4.28)$$

From the general discussion of isotropic metrics above, one still has isotropic dielectric strength and magnetic permeability tensors, but now:

---

([1]) Although that effect is quite small for typical celestial objects, nevertheless, the GPS satellites have to correct for the fact that they are in a weaker part of the Earth's gravitational field than the users that are close to the surface of the Earth or else the position estimates would include a noticeable error.

$$\varepsilon(t, r) = X(t, r)^{-2} \varepsilon_0, \quad \mu(t, r) = X(t, r)^{-2} \mu_0. \qquad (4.29)$$

Once again, the impedance will have the classical value:

$$Z = \frac{\varepsilon_0}{\mu_0}, \qquad (4.30)$$

while the speed of light and index of refraction will take the form:

$$c(t, r) = X(t, r)^2 c_0, \qquad n(t, r) = X(t, r)^{-2}. \qquad (4.31)$$

However, without knowing more specific details about the nature of the function $X(t, r)$, one cannot say much more than that about the last two parameters.

*f. The Kerr metric.* The Kerr metric is yet another generalization of the Schwarzschild metric to the case of a gravitating mass distribution that is rotating. Hence, spherical symmetry is replaced with axial symmetry. If the rotating massive body has mass $M$ and angular momentum $Ma$ (so $a = \omega k^2$, where $\omega$ is the angular velocity and $k$ is the radius of gyration) then its metric will take the form:

$$ds^2 = - c_0^2 \frac{\Delta - a^2 \sin^2\theta}{\Sigma} dt^2 - \frac{4Mar\sin^2\theta}{\Sigma} dt\, d\phi + \frac{\Sigma}{\Delta} dr^2$$
$$+ \left[\frac{(r^2+a^2)^2 - \Delta a \sin^2\theta}{\Sigma}\right] \sin^2\theta\, d\phi^2 + \Sigma\, d\theta^2, \qquad (4.32)$$

in which:

$$\Sigma \equiv r^2 + a^2 \cos^2\theta, \qquad \Delta \equiv r^2 + a^2 - 2Mr. \qquad (4.33)$$

This spacetime then gives us an example of a metric that is not in time + space form, due to the presence of the non-zero component $g_{0\phi}$; another example is the Gödel universe, which is also rotating, although in a somewhat unphysical, rigid way. The effect of the non-vanishing time-space component of $g$ is to make the matrix $\alpha^i_j$ non-vanishing, now. In particular, from (3.31):

$$\alpha^{ij} = - \mu_0\, \varepsilon^{ijk} g_{0k}, \qquad (4.34)$$

so the non-zero components of $\alpha^{ij}$ will be:

$$\alpha^{r\phi} = - \alpha^{\phi r} = - \mu_0\, g_{0\phi} = \frac{4\mu_0 Mar\sin^2\theta}{\Sigma}. \qquad (4.35)$$

One expects that this fact might be associated with some optical effect that might be analogous to Faraday rotation or optical activity. Indeed, Plebanski [**6**] has shown that, in theory, the rotation of the gravitating mass should lead to a rotation of the plane of polarization of the light wave.

The main consequence of the Kerr solution that has been established experimentally is the *Lense-Thirring effect* [21], which says that the rotation of the source distribution will be accompanied by a "dragging of the frames" in its vicinity, in such a way that angular momentum vectors, such as for gyroscopes, will precess. That might relate to the aforementioned rotation of the plane of polarization if one considers that circularly-polarized photons have intrinsic angular momentum in the form of helicity, so the precession of that helicity would be consistent with the rotation of the plane of polarization.

Another variant of the Kerr metric describes a charged, rotating, spherically-symmetric mass distribution. That solution is often proposed as a model for a general-relativistic, but non-quantum, spinning electron that would be analogous to the Dirac electron.

*g. Gravitational fields at large distances.* In Plebanski's proof of the rotation of the plane of polarization, he used the approximation for the gravitational field of a rotating mass that would be valid at large distances from the mass and was given in § 105 of Landau and Lifschitz [20]. At large distances, gravitating masses are usually regarded as point-like, and thus, spherically-symmetric, and if one considers only the static case then one is basically looking at an approximation to the Kerr metric.

If the angular momentum of the rotating mass $M$ is represented by the anti-symmetric matrix $J_{ij} = \varepsilon_{ijk} J^k$ and the unit vector in the radial direction $\hat{\mathbf{r}}$ has components $\hat{x}^i = x^i / r$ then we shall define:

$$\mathfrak{j}_i = J_{ij}\hat{x}^j = \varepsilon_{ijk}\hat{x}^j J^k = (\hat{\mathbf{x}} \times \mathbf{J})_i \,. \tag{4.36}$$

The components of the approximate metric will then take the form:

$$g_{00} = -\left(1 - \frac{2k}{c^2}\frac{M}{r}\right), \qquad g_{0i} = -\frac{2k}{c^3}\frac{\mathfrak{j}_i}{r^2}, \qquad g_{ij} = \delta_{ij}\left(1 - \frac{2k}{c^2}\frac{M}{r}\right). \tag{4.37}$$

In order to invert the metric $g_{\mu\nu}$, we first define the expression:

$$\mathfrak{g}^2 = g_{00}^2 + g_{0i}\, g_{0i} = \left(1 - \frac{2k}{c^2}\frac{M}{r}\right)^2 + \left(\frac{2k}{c^3 r^2}\right)^2 \|\mathfrak{j}\|^2 \,, \tag{4.38}$$

which is the square of the four-dimensional Euclidian norm of the first row (or column) of the matrix of $g_{\mu\nu}$. One notes that when the angular momentum of the gravitating mass vanishes, $\mathfrak{g}^2$ will become simply $g_{00}^2$.

With that definition, we will have:

$$g^{00} = \frac{g_{00}}{\mathfrak{g}^2} = -\frac{1}{\mathfrak{g}^2}\left(1 - \frac{2k}{c^2}\frac{M}{r}\right), \tag{4.39}$$

$$g^{0i} = \frac{g_{0i}}{\mathfrak{g}^2} = -\frac{2k}{\mathfrak{g}^2 c^3} \frac{\mathfrak{j}_i}{r^2}, \tag{4.40}$$

$$g^{ij} = -\frac{1}{g_{00}} \left( \delta_{ij} - \frac{g_{0i} g_{0j}}{\mathfrak{g}^2} \right) = \frac{\mathfrak{g}^2}{\left(1 - \frac{2kM}{c^2 r}\right)} \left[ \delta_{ij} - \left(\frac{2k}{c^3 r^2}\right) \frac{\mathfrak{j}_i \mathfrak{j}_j}{\mathfrak{g}^2} \right]. \tag{4.41}$$

As a check, we see that in the limit of vanishing $\mathfrak{j}_i$, the inverse metric will become:

$$g^{00} = \frac{1}{g_{00}} = -\left(1 - \frac{2k}{c^2} \frac{M}{r}\right)^{-1}, \quad g^{0i} = 0, \quad g^{ij} = -\frac{1}{g_{00}} \delta_{ij} = \left(1 - \frac{2k}{c^2} \frac{M}{r}\right)^{-1} \delta_{ij}, \tag{4.42}$$

which is consistent with the inversion of (4.37) in the same limit.

The associated constitutive matrices will then be:

$$\varepsilon^{ij} = \frac{1}{\mu_0} \left[ \delta_{ij} - \left(\frac{2k}{c^3 r^2}\right) \frac{\mathfrak{j}_i \mathfrak{j}_j}{\mathfrak{g}^2} \right], \quad \alpha_{ij} = \frac{2k}{\mu_0 c^3 r^2 \mathfrak{g}^2} \varepsilon_{ijk} \mathfrak{j}_k, \quad \tilde{\mu}_{ij} = -\frac{1}{\mu_0} \left(1 - \frac{2kM}{c^2 r}\right) \delta_{ij}, \tag{4.43}$$

which makes the impedance:

$$Z = \frac{1}{\mu_0 \mathfrak{g}^2} \left(1 - \frac{2kM}{c^2 r}\right). \tag{4.44}$$

Determining the principal indices of refraction, or dually, the principle speeds of light propagation would involve diagonalizing the metric tensor, which we shall pass over here in the interests of brevity.

*h. Bianchi type I metrics.* The Bianchi type I metrics allow one to deal with anisotropic, but homogeneous, metrics in full generality. In Cartesian coordinates, they take the generic form:

$$ds^2 = -c_0^2 \, dt^2 + n_1(t)^2 \, dx^2 + n_2(t)^2 \, dy^2 + n_3(t)^2 \, dz^2, \tag{4.45}$$

for suitable functions of time $n_i(t)$, $i = 1, 2, 3$.

Hence, one will have:

$$g_{00} = -c_0^2, \quad g_{ij} = \begin{bmatrix} n_1(t)^2 & 0 & 0 \\ 0 & n_2(t)^2 & 0 \\ 0 & 0 & n_3(t)^2 \end{bmatrix}, \tag{4.46}$$

$$g^{00} = -c_0^{-2}, \quad g^{ij} = \begin{bmatrix} n_1(t)^{-2} & 0 & 0 \\ 0 & n_2(t)^{-2} & 0 \\ 0 & 0 & n_3(t)^{-2} \end{bmatrix}. \tag{4.47}$$

This will make:

$$\varepsilon^{ij} = \varepsilon_0 \begin{bmatrix} n_1(t)^2 & 0 & 0 \\ 0 & n_2(t)^2 & 0 \\ 0 & 0 & n_3(t)^2 \end{bmatrix}, \quad \mu^{ij} = \mu_0 \begin{bmatrix} n_1(t)^{-2} & 0 & 0 \\ 0 & n_2(t)^{-2} & 0 \\ 0 & 0 & n_3(t)^{-2} \end{bmatrix}. \quad (4.48)$$

This suggests that the principal axes of both tensors must coincide with the Cartesian axes in question.

Once again, the spatial impedance will retain its classical value:

$$Z = \varepsilon_0 / \mu_0. \quad (4.49)$$

In order to discuss the speed of light in such a medium, we note that if the metric components $n_i(t)^2$ can be treated as principal indices of refraction when the Cartesian axes are defined by a principal frame then one might set:

$$n_i(t) = \frac{c_0}{c_i(t)}, \quad i = 1, 2, 3, \quad (4.50)$$

which would make the speeds of propagation in the principal directions equal to:

$$c_i(t) = \frac{c_0}{n_i(t)}, \quad i = 1, 2, 3. \quad (4.51)$$

However, since $|n_i(t)| \leq c_0$ for any real-world index of refraction, that would have to be a necessary condition for a metric to admit such an interpretation of its spatial components.

If such an interpretation is possible then since one might also set:

$$(n_i)^2 = \frac{\overline{\varepsilon}_i \overline{\mu}_i}{\varepsilon_0 \mu_0} \quad \text{(no sum)} \quad i = 1, 2, 3, \quad (4.52)$$

the tensors $\varepsilon^{ij}$ and $\mu^{ij}$ above can be put into the form:

$$\varepsilon^{ij} = \begin{bmatrix} \dfrac{\overline{\varepsilon}_1 \overline{\mu}_1}{\mu_0} & 0 & 0 \\ 0 & \dfrac{\overline{\varepsilon}_2 \overline{\mu}_2}{\mu_0} & 0 \\ 0 & 0 & \dfrac{\overline{\varepsilon}_3 \overline{\mu}_3}{\mu_0} \end{bmatrix}, \quad \mu^{ij} = \begin{bmatrix} \dfrac{\overline{\varepsilon}_1 \overline{\mu}_1}{\varepsilon_0} & 0 & 0 \\ 0 & \dfrac{\overline{\varepsilon}_2 \overline{\mu}_2}{\varepsilon_0} & 0 \\ 0 & 0 & \dfrac{\overline{\varepsilon}_3 \overline{\mu}_3}{\varepsilon_0} \end{bmatrix}, \quad (4.53)$$

and if one defines the principal values of the tensors $\varepsilon^{ij}$ and $\mu^{ij}$ to be $\varepsilon_i$ and $\mu_i$ ($i = 1, 2, 3$), respectively, then that would make:

$$\varepsilon_i = \frac{\bar{\mu}_i}{\mu_0}\bar{\varepsilon}_i, \qquad \mu_i = \frac{\bar{\varepsilon}_i}{\varepsilon_0}\bar{\mu}_i. \tag{4.54}$$

**5. Effective metrics.** As mentioned above, in 1923, Gordon [**6**] pointed out that the electromagnetic constitutive properties of a material medium that is in a state of relative motion to an observer can alter the propagation of light in that medium in such a way the light cones of the background space-time will be deformed into light cones of an effective metric that depends upon the constitutive properties and the relative velocity.

Actually, the scope of that association was confined to linear, isotropic, non-dispersive media whose constitutive properties could be characterized by a dielectric strength $\varepsilon(t, x)$ and a magnetic permeability $\mu(t, x)$. If $u = u_\mu \, dx^\mu$ is the relative covelocity 1-form (with respect to the background metric $g$) of the medium then the effective metric that Gordon defined took the form:

$$\gamma = g + \left(1 - \frac{1}{\varepsilon\mu}\right) u \otimes u; \tag{5.1}$$

i.e., in component form:

$$\gamma_{\mu\nu} = g_{\mu\nu} + \left(1 - \frac{1}{\varepsilon\mu}\right) u_\mu u_\nu. \tag{5.2}$$

The individual components of the effective metric $\gamma$ are then:

$$\gamma_{00} = g_{00} + \left(1 - \frac{1}{\varepsilon\mu}\right)(u_0)^2, \quad \gamma_{0i} = g_{0i} + \left(1 - \frac{1}{\varepsilon\mu}\right) u_0 u_i, \quad \gamma_{ij} = g_{ij} + \left(1 - \frac{1}{\varepsilon\mu}\right) u_i u_j. \tag{5.3}$$

For Gordon, the relationship between the field strength 2-form $F$ and the excitation bivector field $\mathfrak{H}$ was then expressed in terms of components as:

$$\mu \mathfrak{H}^{\mu\nu} = \sqrt{-g} \, \gamma^{\mu\kappa} \gamma^{\mu\lambda} F_{\kappa\lambda}. \tag{5.4}$$

This clearly has the form of a linear, non-dispersive constitutive law. For us, the factor of $\sqrt{-g}$ is misplaced, since it properly belongs with the component of the Lorentz-invariant volume element on space-time:

$$V = \sqrt{-g} \, dx^0 \wedge dx^1 \wedge dx^2 \wedge dx^3 = \frac{1}{4!}\sqrt{-g} \, \varepsilon_{\kappa\lambda\mu\nu} \, dx^\kappa \wedge dx^\lambda \wedge dx^\mu \wedge dx^\nu. \tag{5.5}$$

It would only get absorbed into the constitutive law when one composes the electromagnetic constitutive isomorphism $C : \Lambda^2 \to \Lambda_2$ with the Poincaré isomorphism # : $\Lambda_2 \to \Lambda^2$ that is defined by the volume element to get a linear isomorphism $\kappa = \# \cdot C : \Lambda_2 \to \Lambda_2$ that reduces to the Hodge duality isomorphism when the constitutive law is simply the raising of indices.

We can see that the present way of associating electromagnetic constitutive properties with an effective metric is more general than the Gordon method, which pertains to only linear, isotropic, and non-dispersive media in a state of (relative) motion. In particular, the present method extends to anisotropic media, as well as some media in which the off-diagonal components are of a more general form than is assumed by the Gordon metric.

**6. Discussion.** As we mentioned above, the scope of the electromagnetic constitutive laws that can be defined by Lorentzian metrics is limited to linear, non-dispersive ones that do not admit any birefringence. However, if one regards the transition from classical electromagnetism to quantum electromagnetism as something that is indexed by the onset of vacuum polarization, and possible vacuum birefringence, and critically-high values of the electric and magnetic field strengths then one must figure that such metrically-defined constitutive laws can only get one so close to that realm of quantum phenomena.

Nevertheless, since the realm of general-relativistic physics, which is mostly astrophysics and cosmology, is more amenable to direct observation than the small neighborhoods of elementary charge distributions, one might hope that one might gain a better intuition of what to expect and what to look for when one goes to those strong field regimes, since general relativity is basically a strong-field theory of gravitation, just as quantum electrodynamics is a strong-field theory of electromagnetism. Furthermore, it is interesting that the recent experimental discovery of gravito-electromagnetism tends to suggest that Maxwell's equations can also serve as the weak-field equations of gravitation, as well as the weak-field equations of electromagnetism. Hence, one hopes that the strong-field equations of gravitation might tell physics what to expect of the strong-field equations of electromagnetism, as well.

### References ([*])

---

([*]) References marked with an asterisk are available in English translation through the author's website: neo-classical-physics.info.

____________